\documentclass[useAMS,usenatbib]{mn2e}
\input psfig.tex

\def\simgt{\stackrel{>}{{}_\sim}}
\def\simlt{\stackrel{<}{{}_\sim}}

\title[$L_X-T$ at $z\simgt1.5$]{Cluster X-ray luminosity--temperature 
relation at $\mathbf{z\simgt1.5}$}

\author[Andreon et al.]{S. Andreon,$^1$\thanks{stefano.andreon@brera.inaf.it},
G. Trinchieri$^1$,
F. Pizzolato$^1$\\
$^1$INAF--Osservatorio Astronomico di Brera, via Brera 28, 20121, Milano, Italy \\
}
\date{Accepted ... Received ...}
\pagerange{\pageref{firstpage}--\pageref{lastpage}}
\pubyear{2010}

\begin{document}
\maketitle

\label{firstpage}

\begin{abstract}
The evolution of the properties of the hot gas that fills the potential
well of galaxy clusters is poorly known, since models are  
unable to give robust predictions and observations lack a
sufficient redshift leverage and are affected by selection effects.
Here, with just two high redshift, $z\approx1.8$, clusters 
avoiding selection biases,
we obtain a significant extension of the redshift
range and we begin to constrain the possible evolution of the
X-ray luminosity vs temperature
relation.  
The two clusters, JKCS\,041 at $z=2.2$
and    
ISCS\,J1438+3414 at $z=1.41$, 
are respectively the most distant cluster
overall, and the second
most distant that can be used for studying scaling relations.
Their location in
the X-ray luminosity vs temperature 
plane, with an X-ray luminosity 5 times
lower
than expected, suggests at the 95\% confidence that
the evolution of  the intracluster medium has not been self--similar 
in the last three quarters of
the Universe age. Our conclusion is reinforced by data 
on a third, X-ray selected, high redshift cluster, 
too faint for its temperature 
when compared to a sample of similarly 
selected objects. 
Our data suggest that non-gravitational effects, such as the baryon physics,
influence
the evolution of galaxy cluster. Precise knowledge of evolution
is central for using galaxy clusters  as cosmological probes in
planned X-ray surveys such as WFXT or JDEM.  
\end{abstract}


\section{Introduction}

The observation of the diffuse, X--ray emitting medium 
(a.k.a. intra-cluster medium,
or ICM) of galaxy clusters provides quantities like its mass,
temperature ($T$) and X--ray luminosity ($L_{X}$).
The analysis of the scaling relation between these physical quantities
gives considerable insight into the physical processes in the ICM (e.g.
Rosati et al. 2002
and reference therein).
On the other hand, the evolution of these scaling relations is difficult to
predict  theoretically (e.g. Norman 2010).
The simplest model
(Kaiser 1986), 
in which the ICM  evolution is governed only by gravity,
predicts an $L_X-T$ relation shallower than observed (Markevich 1998).
This suggests that non-gravitational energy inputs, such as merger shocks
or feedback from active galactic nuclei (AGNs) 
and star formation, need to be considered.
More sophisticated models sensitively depend on the assumed physics 
of the baryons, and their
predictions can be tuned to be in good agreement with 
observed scaling relations
(Kravtov et al. 2005; Nagai, Kratsov \& Vikhlinin 2007; Bode, Ostriker \&
Vikhlinin 2009) 
measured in the nearby
Universe, if one accept an overprediction of the baryon fraction in stars 
by an order of magnitude (Gonzalez, Zarisky \& Zabludoff 2007, Andreon 2010).

The most direct way to probe ICM evolution is to measure  the 
scaling relations  over a  wide  range of redshifts.  Here a difficulty
arises: many cluster samples with known $L_X$ and $T$ are either 
X-ray selected, or are
heterogenous collections of objects without a simple and accountable
selection function. In both cases, neglecting the 
selection function may bias the $L_X-T$ relation (Stanek et al. 2006;
Pacaud et al. 2007; Nord et al. 2008),
because at a given temperature clusters more luminous 
enter more easely in the sample (they can be seen on a larger volume, have
smaller temperature errors, and are more frequently in archive and samples).
Therefore, the mean $L_X$ at a given $T$ can be systematically
over-estimated,
unless one accounts for the selection function 
(e.g. Gelman et al. 2004, Pacaud et al. 2007, Andreon \& Hurn 2010).
The requirement of a known selection function restricts the choice of
the available samples and the redshift baseline  making 
hard to detect deviations from a self--similar evolution 
for lack of extension at high redshift. For example, $z\le 1.05$ 
for 
Pacaud et al. (2007), 
and $z<0.2$ for 
Pratt et al. (2009).

Only a handful of clusters are known at high $z$ (four at $z>1.4$).
In this paper we use the only two suitable for this study, namely 
JKCS\,041, %
probably the most distant cluster known to date, 
and ISCS\,J1438+3414 (at $z=1.41$, 
Stanford et al., 2005),
the second most distant cluster that can be used for studying scaling relations.
Note that the redshift of JKCS\,041, conservatively
estimated at $z=1.9$ in Andreon et al. (2009) and has now a
red-sequence estimated redshift of $z=2.20\pm0.11$ 
(Andreon \& Huertas-Company 2010).
Both are optically-NIR selected, i.e. are detected through their
galaxies, and have been subsequently  followed up in X rays  (see 
Andreon et al. 2009 
for JKCS\,041 and this paper for ISCS\,J1438+3414) to derive $L_X$ and T for the gas. Though 
small, this sample is free from the biases that affect X-ray selected
samples, since these clusters are considered independently from
their X--ray luminosity.
By using them, we extend the redshift
baseline to $z\sim 2$, where the self-similar model predicts
a brightening  1.7 times larger than at $z=1$.

\begin{figure}
\centerline{%
\psfig{figure=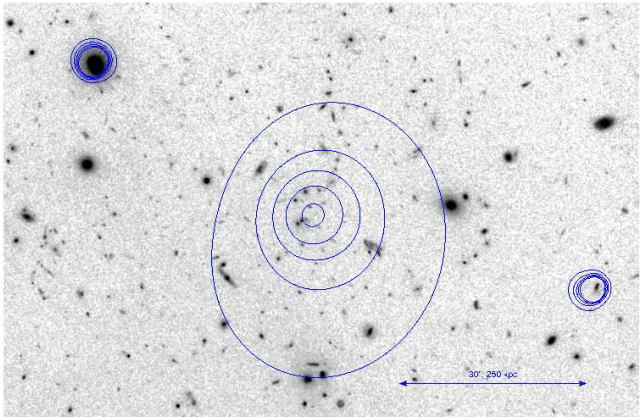,width=8.5truecm,clip=}%
}
\caption[h]{Contours from an adaptively smoothed Chandra image in the
[0.3-2] keV energy band superposed onto an Hubble Space Telescope  
(F850LP) image of ISCS\,J1438+3414.}
\end{figure}

We adopt the following cosmological parameters:
$\Omega_\Lambda=0.7$, $\Omega_m=0.3$ and $H_0=70$ km s$^{-1}$
Mpc$^{-1}$.
The scale, at $z=1.41$, is 8.4 kpc arcsec$^{-1}$.
As point estimate and
error measurements, we quote 
posterior mean and standard deviation when
a Bayesian approch is esplicitely mentioned, or, otherwise the usual
profile likelihood-based estimates (e.g. XSPEC error, 
$-2 \Delta \ln \mathcal{L} = 1$).

\section{Data and Analysis}

\subsection{HST Observations}

ISCS\,J1438+3414 has been observed with the Wide Field Camera of the Advanced Camera
for Surveys (hereafter ACS, Ford et al. 1998, 2002)
of the {\it Hubble Space Telescope} (HST, hereafter) for 10 ks with the F850LP
filter. These data
are reduced following the procedure adopted in, e.g., 
Andreon (2008): 
the raw ACS data 
were processed through the standard CALACS pipeline
(Hack 1999)
at STScI. This includes overscan, bias, and dark subtraction, as well as
flat-fielding. Image
combination has been done with the multidrizzle software
(Koekemoer et al. 2002). 
The data quality arrays enable
masking of known hot pixels and bad columns, while cosmic rays and other
anomalies are rejected through the iterative drizzle/blot technique. 
Fig 1 shows the resulting image. 

\begin{figure}
\centerline{
\psfig{figure=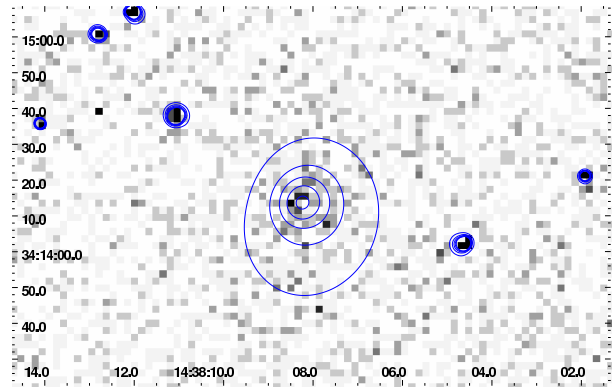,width=8truecm,clip=}}%
\caption[h]{
[0.3-2] keV Chandra X-ray image of ISCS\,J1438+3414, binned
to 2 arcsec pixels. 
The image is overlaid with contours of the X-ray emission after adaptive
smoothing so that all features are significant to at least the 3 $\sigma$ level. 
The faintest contour was chosen to closely approximate the region where 
the smoothing kernel contained a signal above the 3
$\sigma$ threshold on a scale of about 20 arcsec.
North is up and East is to the left.
}
\end{figure}

\subsection{Chandra X-ray Observations}

ISCS\,J1438+3414 was observed by Chandra for $150$ ks on 2009 October 4 and 9 (ObsID 
10461 \& 12003), using the ACIS-S detector. The data were reduced using  the standard
data reduction procedures and were checked for periods of high background.  We
found no differences between the data quality and the set-up of the instruments of the
first and second observation.  We therefore merged the two datasets for a total obsering
time of 143 ks, consistent with the 150 ks originally requested.
A preliminary examination of the data showed 
that the $0.3-2.0$ keV energy band gave the maximum cluster signal to
noise ratio for our image analysis. 
The image produced in this energy band is shown in
Fig. 2. The image was then adaptively smoothed
with the Ebeling et al.
(2006) algorithm, avaiable in the CIAO software, 
requesting a minimum significance of $3\sigma$.  Contours of this smoothed
X-ray image are overlaid in Fig. 1 on the HST F850LP image.  
The X-ray morphology
appears regular, but this could simply result from the relatively large
kernel required by the low signal-to-noise  of the cluster emission ($\sigma
\simlt 20''$).
Within a 1 arcmin radius from the cluster centre
there are $274 \pm 60$ photons in the 0.3-2 keV band (after subtraction of the
background and exclusion of point sources).

\subsubsection{X-ray Image Analysis}

To quantify the cluster surface
brightness distribution,
the Chandra image of ISCS\,J1438+3414 was fit with a two-dimensional (2D)  
beta profile (Cavaliere \& Fusco-Femiano 1978) 
with an additive 
constant  (on detector) component for the background\footnote{ We checked that  
consistent results are found whether we model the background with
a constant on the detector or we modulate it through the telescope vignetting.}. 
The model was constrained to be circular. 
Point sources were masked out 
during the fitting process. We
adopt the Bayesian approach of 
Andreon et al. (2008) 
with uniform priors
except for $\beta$, taken to be a
Gaussian, zero-ed at $3 \beta - 1/2 <1$ (the beta model must have
a finite integral), centered on $\beta=2/3$ and with width $\sigma_\beta =0.2$,
the latter to account for the fact that clusters tend to 
have $\beta \approx 2/3$ (e.g. Maughan et al. 2008).
The posterior probability distribution of $\beta$ values 
resulting from the fit  is displayed in Fig 3 and compared to
the assumed prior. 
We found $3 \beta -1/2 = 1.2 \pm 0.15$, but with a posterior
distribution fairly different from a Gaussian (see Fig 3),
implying that the data carry information about the beta parameter. Basically,
the data constrain $\beta$ to be small, $\beta \simlt 2/3$ at 95 \% confidence
(with $\beta >1/2$ to ensure a finite flux), but not its exact value.

\begin{figure}
\centerline{\psfig{figure=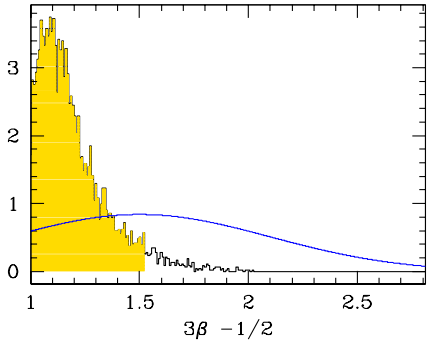,width=6truecm,angle=0}}
\caption[h]{Posterior probability distribution for $3 \beta -1/2$.
The black jagged histogram shows the posterior, marginalised 
over the other parameters.  The jagging is due to the finite lenght
of the chain sampled, i.e. is noise, not signal.
The shaded (yellow) range shows
the 95\% highest posterior credible interval. The blue smooth
curve shows the assumed prior for the parameter. The data constrain
$3 \beta -1/2$ to be small (e.g. $\beta<2/3$ at 95 \% confidence).
}
\end{figure}

Figure 4 shows an azimuthally averaged 
radial profile of the data with euristic 
error bars (for visualisation purposes) and the
mean 2D model, with 68 \% (highest posterior) error (shaded).
The latter rigorously accounts for uncertainty and co-variance of all modelled
quantities.  We emphasize that the model was
not fit in this space. The X-ray emission is manifestly extended with respect to
the Chandra 0.5 arcsec point spread function. 
The fit coordinates of the X-ray emission
of ISCS\,J1438+3414 are 
$RA = 14:38:08$ $\pm3$ arcsec and $DEC = +34:14:14$ $\pm3$
arcsec. 
We found a core radius of $9\pm2$ arcsec 
($75$ kpc). We also compute the core radius with $\beta$ fixed at $2/3$,
for comparison with other clusters, r$_c \sim 12\pm2$ arcsec ($100$ kpc).
In either case, r$_c$ is in the range of values observed for local clusters.

\begin{figure}
\centerline{
\psfig{figure=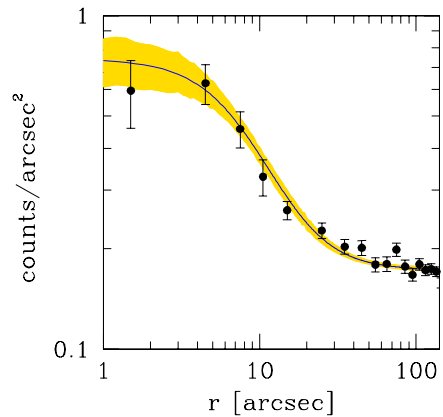,width=6truecm,angle=0}}
\caption[h]{
Radial profile of ISCS\,J1438+3414.
The solid line marks the mean beta model. The shaded region marks the 68\%
highest posterior credible interval for the model. 
Error bars on the data points are euristically computed, and,
do not account, for example, for the
intensity gradient across the bin, the uncertainty on the
center, etc. The shading, instead, does. 
This figure is simply for visualisation purposes, 
the model was not fit in this space.
}
\end{figure}

\subsubsection{X-ray Spectral Analysis}

Our spectral analysis procedure was chosen to match that of 
Pacaud et al. (2007) 
to allow direct comparison with their $L_X-T$ relation. In
summary, a cluster spectrum was extracted from an aperture of radius
$30$ arcsec (252 kpc) (with minor masking of a single point source falling just
outside the boundary), chosen to maximise the
signal to noise ratio. A background spectrum was extracted from two
regions around the cluster, sufficiently separated to exclude
any cluster emission (mean backgroud radius: 125 arcsec, 1050 kpc at the
cluster distance), and chosen to be included in the same chip,
but avoiding gaps and bad columns. The resulting cluster spectrum containes
$\sim 280$ net photons in the $0.3-7.0$ keV band used for spectral
fitting. The source
spectrum was fit with an absorbed APEC
(Smith et al. 2001)
plasma model, with the absorbing column fixed at the Galactic value
($0.98\times10^{20}$ cm$^{-2}$, Dickey \& Lockman 1990),   
the metal abundance fixed at $0.3$ relative to Solar
and the redshift of the plasma model fixed at $1.41$. 
The spectrum was grouped to contain a minimum of $5$ counts per
bin and the source and background data were fit within the 
XSPEC spectral package using the
modified C-statistic (also called W-statistic in XSPEC).
Simulations in Willis et al. (2005) 
confirm that this methodology is reliable.

The best fitting spectral model (plotted in Fig. 5) gives 
$kT=4.9^{+3.4}_{-1.6}$ keV, which results in an unabsorbed
bolometric X-ray flux of $1.4\times10^{-14}$ erg cm$^{-2}$ s$^{-1}$. 

In order to measure X-ray scaling relations we need $L_X(<r_{500})$, and
therefore we need to estimate $r_{500}$,
which is derived from the cluster temperature,  
using the scaling relation of 
Finoguenov et al (2001) 
as given in equation 2 of 
Pacaud et al (2007). 
For the best fit
temperature, $r_{500}$ $=0.48$ Mpc, but temperature has errors, which
we need to account for.
We use a Bayesian approach: 
for each temperature 
(we used a chain of 1000 samples drawn from the temperature likelihood)
we compute 1000 estimated values of $r_{500}$. For each $r_{500}$ and for
each sampling the posterior distribution of
the parameters of the $\beta$ model (a chain of 2000 values),
we compute the ratio between the flux in the spectral aperture
and within the estimated $r_{500}$, including
correction for point sources. This gives the wanted posterior
distribution of the conversion factor. It turns out to have
a (near to) log-normal shape, i.e. it is normal after
moving to log units. We found $\log c = 0.16 \pm0.06$ dex, i.e.
the conversion factor has a $14$ \% uncertainty. This uncertainty
is larger than the uncertainty on the flux in the spectral
aperture alone (10 \%), and therefore cannot be neglected.
Not accounting for the temperature error also induces a bias
almost as large as the flux error in the spectral aperture.
To summarize,
the bolometric luminosity within
$r_{500}$ is  $L_X (<500) = (2.5\pm 0.5)\times10^{44}$ erg s$^{-1}$.
 We emphasize that this is the luminosity within the
angular aperture of radius $r_{500}$.

\begin{figure}
\centerline{\psfig{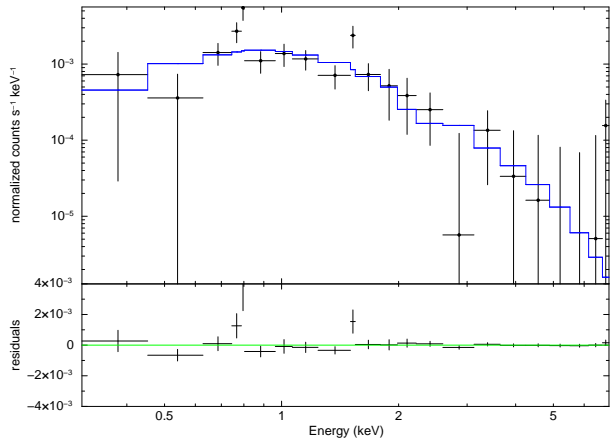}}
\caption[h]{The Chandra X-ray spectrum and best fitting model of
ISCS\,J1438+3414 are 
shown in the top panel, with the residuals shown in the bottom panel. 
The spectrum is rebinned for displaying purposes, but is  fitted
on a minimally binned version.}
\end{figure}

Finally, the temperature of ISCS\,J1438+3414 can be used to estimate the cluster's
mass.  Under the (strong)
assumption that the temperature mass 
relation presented in Finoguenov et al. (2001) 
self-similarly evolves (doubtful, but
adopted for lack of anything more suitable)
from $z=0$ to $z=1.4$, and neglecting all (at this point negligible)
statistical subtilities, we found 
$M_{500}=2.0^{+2.6}_{-0.9}\times10^{14}M_\odot$\footnote{Although
not as clearly stated as in this work, 
the mass of JKCS\,041 quoted in Andreon et al. (2009) has been also
derived self-similarly evolving the relation.}.

\section{A first look at $\mathbf{L_X-T}$ scaling relation at 
$\mathbf{z\approx 1.8}$}

Figure 6 shows the position of the two clusters in the 
X-ray luminosity, $L_X(<r_{500})$, vs X-ray 
temperature $T$ plane relative to the $L_X-T$ relation self-similarly 
evolved at the redshift of the two clusters.  
Because of the slighly revised redshift from the publication of 
Andreon et al. (2009), JKCS\,041 data have been re-analized with
the updated redshift. We find:
$L_X (<500) = (9.1\pm 2.5)\times10^{44}$ erg s$^{-1}$ and 
$kT=7.3^{+6.7}_{-2.6}$ keV. Once the large temperature errors
have been taken into account, it is plausible to find a cluster such as
JKCS\,041 in the volume surveyed in Andreon et al. (2009)
in a standard $\Lambda$CDM universe.

The relation is derived from data presented in 
Pacaud et al. (2007). 
In their paper, the authors 
account for the  
selection function, but 
did not publish the value of the parameters of the 
$L_X-T$ scaling.  We obtained 
the selection function in electronic form directly from the authors, 
through a Bayesian analysis 
(Andreon \& Hurn 2010)
we recomputed the $L_X-T$ scaling at the median redshift 
of their
sample, $z=0.33$,
and we checked that our results is entirely consistent with theirs.
The scatter amplitude uses as prior the
Stanek et al. (2006) 
measurements.
The relation, self--similarly evolved at $z=1.41$ (solid blue line) and $z=2.2$
(dashed red line), is shown in Fig.~6. 

\begin{figure}
\centerline{\psfig{figure=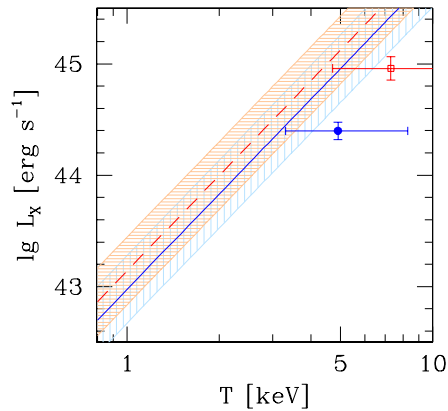,width=6truecm,clip=}}
\caption[h]{Location of ISCS\,J1438+3414 a $z=1.41$ (lower closed point), and
JKCS\,041 at $z\sim 2.2$ (upper open point) in the $L_X-T$ plane. The blue solid (red dashed) 
line marks the $z=0.33$ $L_X-T$ scaling relation self-similarly evolved 
to $z=1.41$ ($z= 2.2$). 
We shaded the region within one intrinsic scatter from the mean model:
the red, horizontal, shading refers to $z=2.2$, whereas the blue, vertical,
shading to $z=1.41$. 
Both clusters are $\sim5 \times$ too faint for their X-ray
temperature if the $L_X-T$ 
scaling relation evolves self-similarly.
}
\end{figure}

Both clusters are located below
the self-similar expectation, too faint by $0.73$~dex
(ISCS\,J1438+3414) and by $0.68$~dex (JKCS\,041), i.e. 
by a factor 5,  for their (best fitting) temperature. On the
other end, they are about only ``1 sigma" away from the predicted scaling
relation, given their relatively large errors on $T$. 
The probability to observe two points ``1 sigma", or more, 
away from the expected
relation and on
the same side is~$5 \%$ ($=0.32*0.32/2$), i.e. our claim is
statistically significant at $95 \%$~confidence
(in the above, $p$-value, sense). 
A more advanced analysis is not very useful:
a) the influence of a redshift uncertainty for JKCS\,041
is negligible: using the previous value of  $z = 1.9$ ($3 \sigma$ away 
from the current value)  makes JKCS\,041
$0.70$~dex too faint (vs $0.68$~dex) and still 1 sigma away from
the predicted scaling relation;
b)
we performed a preliminary account for the fact that
points are not exaclty ``1 sigma" away and for the covariance between 
regressed quantities
($T$, on abscissa, enters also in the ordinate, via $r_{500}$), but the ultimate
limit is given by the sample size, not by the precise treatment of errors, and 
to improve the former more data are needed, not a better statistical analysis.

We have not included in our analysis the only remaining cluster at 
$z>1.4$ for which a  measure of $L_X$ and $T$ is avalable, namely
XMMXCSJ2215.9-1738 at $z=1.46$ 
(Stanford et al. 2006) 
because this cluster is X-ray selected and its (X-ray) 
selection function unpublished. As already 
noted by Hilton et al. (2010),
this high redshift cluster is 
too faint for its temperature 
when compared to a sample of similarly selected objects from
Maughan et al. (2006) 
and when the selection function is ignored. 
If we assume that the X-ray selection 
factors out (i.e. it is benign), 
our suggestion of a breaking of the self-similar 
evolution is reinforced and its statistical significance 
increased.

\section{Conclusions \& Discussion}
 
The large redshift leverage 
considered in this paper has provided a direct, though not yet compelling, 
evidence that clusters do not evolve self-similarly 
in the last $10.6$~Gyr, about three quarter of the current age of the Universe. 
We remark that our result relies on a  large  redshift 
leverage, rather than on a detailed analysis of  small effects 
on  large samples at lower redshift. 
If confirmed, the trend we have found  implies that non-gravitational
effects, such as baryon physics, %
began long ago to shape the clusters' scaling relations.
In particular, the observed evolution is in line with the 
predictions of simulation that include high-redshift
pre-heating and radiative cooling in addition 
to shock heating, such as those in Short et al.
(2010). They predict that our clusters should be a factor 3 to 4 fainter
than self--similar evolution while we observe a factor 5. 
Instead, their models that include
feedback directly tied to galaxy formation or that
incorporate gravitional heating only
strongly disagree with our observations. 
This conclusion should not
over-emphasized, because we are still a long away from having
the numerical resolution required to really implement these mechanisms
(e.g. Norman 2010), for example to follow the formation of stars, whose
feedback is deemed important for the evolution of the gas properties.

It is of the utmost importance to extend the sample of 
non X-ray selected clusters 
to $z> 1.4$, to confirm the modulation provided
by non-gravitational phenomena in the cluster evolution. 
We emphasize the need of non-X-ray selected samples:  X-ray 
selected samples should be treated with caution when used in this 
context, because the probability that an object is in the sample is not random
in $L_X$ at a given $T$. Optically/near-infrared selected samples should instead 
be used since their selection is not
due to their X-ray properties, unless we were able to predict their individual 
X-ray luminosity relative to the average X-ray luminosity at
a given $T$ in absence of X-ray data and we were to make
use of this information to select the objects.

If confirmed, the breakdown of the self--similar evolution,
would have important consequences for the cosmological studies.
Indeed, the evolution with redshift of the scaling relations is very
sensitive to cosmological parameters (e.g. 
Allen et al. 2004; Albrecht et al. 2006, Report of the Dark Energy Task Force, 
and references therein). 
A proper assessment of the intrinsic processes shaping the 
scaling relations is fundamental for the use of
galaxy cluster surveys, such as the planned WFXT 
(Conconi et al. 2010) and JDEM (Sholl et al. 2009), 
as probes of the cosmological parameters.

\section*{Acknowledgements}

We thank Ben Maughan for his encouragement at the start of this work,
Florian Pacaud for giving us the selection function of his sample
in electronic form, and the anonymous referee for useful suggestions.
This paper is based on observations obtained by 
Chandra (ObsID 10461 \& 12003) and HST (10496).
We acknowledge financial contribution from the agreement ASI-INAF I/009/10/0.

{}

\bsp

\label{lastpage}

\end{document}